\documentclass[twocolumn,prl,amsmath,showpacs]{revtex4}
\usepackage{graphics}
\newcommand{\be}{\begin{equation}}
\newcommand{\ee}{\end{equation}}
\newcommand{\ber}{\begin{subequations}\begin{eqnarray}}
\newcommand{\eer}{\end{eqnarray}\end{subequations}}
\newcommand{\berb}{\begin{eqnarray*}}
\newcommand{\eerb}{\end{eqnarray*}}
\newcommand{\av}[1]{\langle #1\rangle_\rho}
\newcommand{\avv}[1]{\langle #1\rangle}

\newcommand{\mO}{{\mathcal O}}

\newcommand{\mU}{{\mathcal U}}
\newcommand{\mI}{{\mathcal I}}

\newcommand{\ket}[1]{\left|#1\right\rangle}
\newcommand{\bra}[1]{\left\langle#1\right|}
\newcommand{\tr}{{\rm Tr}}
\begin{document}
\title{Entanglement Detection by Local Orthogonal Observables}
\author{Sixia Yu, Nai-le Liu}
\affiliation{Hefei National Laboratory for Physical Sciences at
Microscale \& Department of Modern Physics, University of Science
and Technology of China, Hefei, Anhui 230026, P.R. China }
\date{\today}

\begin{abstract}
We propose a family of entanglement witnesses and corresponding
positive maps that are not completely positive based on local
orthogonal observables. As applications the entanglement witness
of a $3\times 3$ bound entangled state [P. Horodecki,
Phys.~Lett.~A {\bf 232}, 333 (1997)] is explicitly constructed and
a family of $d\times d$ bound entangled states is introduced,
whose entanglement can be detected by permuting local orthogonal
observables. The proposed criterion of separability can be
physically realized by measuring a Hermitian correlation matrix of
local orthogonal observables. \pacs{03.67.Mn, 03.65.Ud, 03.65.Ta}
\end{abstract}
\maketitle

{\it Introduction.---}Entangled states are valuable resources for
the quantum computation and communication. However the boundary
between the entangled states and the separable states, states that
can be prepared by means of local operations and classical
communications \cite{werner}, is still not well characterized.
Entanglement detection turns out to be a rather tantalizing
problem.

There have been many approaches to the problem such as the partial
transposition criterion \cite{peres, horo1}, the realignment
criterion \cite{chen}, the symmetric extension criterion\cite{sym,
symm}, and the equation-solving method \cite{wu}, to name a few.
Many criteria such as the partial transposition criterion and the
reduction criterion arise from positive maps that are not
completely positive (non-CP). A state is separable if and only if
the state keeps its positivity under all non-CP maps \cite{horo1}.
The states with positive partial transposition (PPT) belong to
bound entangled states \cite{be} while the states violating the
reduction criterion can be distilled, or free entangled
\cite{reduct}. The non-CP maps are not very easy to find and they
are not physically realizable. There are also some physical
approaches including Bell inequalities \cite{bell,terha,yu1},
local uncertainty relationships \cite{yu2,hofman,guh}, and
entanglement witnesses \cite{terha,reduct}. A 3-setting Bell like
inequality is found to be a sufficient and necessary condition for
the $2\times2$ system \cite{yu2}. A local uncertainty relation is
found to be violated by bound entangled states \cite{hofman,guh}.

 In this Letter we shall at first construct a family of entanglement
 witnesses, from which
 a generalization of the reduction criterion can be
 derived, based on local orthogonal observables.
 Then we apply our criterion of separability to
 several bound entangled states,  including
 a family of bound entangled states where the criterion is sufficient and
 necessary. Finally we reformulate the criterion in terms of
 physically measurable quantities, namely Hermitian correlation matrices.

{\it Local orthogonal observables.---}We consider a $d\times d$
system, a bipartite system with two $d$-level subsystems labelled
by $A$ and $B$, whose Hilbert space is spanned by $\ket{m,n}=\ket
m\otimes\ket n, (m,n=1,2,\ldots, d)$. For each system a complete
set of {\it local orthogonal observables} (LOOs) is a set of $d^2$
observables $A_\mu$ $(\mu=1,2,\ldots,d^2)$ of this system
satisfying orthogonal relations
 \be\tr A_\mu A_\nu=\delta_{\mu\nu},\quad
 (\mu,\nu=1,2,\ldots,d^2).
 \ee
The set of LOOs is {\it complete} in two senses. Firstly they form
an orthonormal base for all the operators in the Hilbert space of
a $d$-level system. For example a density matrix $\varrho_A$ may
have an expansion $\varrho_A=\sum_\mu\tr(\varrho_AA_\mu)A_\mu$.
Secondly $d^2$ states $\ket{A_\mu}\equiv A_\mu\ket\Phi$ form an
orthonormal basis for the composite system, where
$\ket\Phi=\sum_i\ket{i,i}$ and $A_\mu$ act on the first subsystem.

In the case of qubits a typical complete set of LOOs can be
$\{I,\sigma_x,\sigma_y,\sigma_z\}/\sqrt2$. For later use, we
define a standard complete set of LOOs
$\{\lambda_\mu\}=\{\lambda_m=\ket m\bra m, \lambda^\pm_{mn}\}$
$(m,n=1,2,\ldots,d)$ where
 \ber
 \lambda^+_{mn}&=&\frac{\ket m\bra n+\ket n\bra
 m}{\sqrt2}\quad(m<n),\\
 \lambda^-_{mn}&=&\frac{\ket m\bra n-\ket n\bra m}{i\sqrt2}
 \quad(m<n).\eer
As testing observables $\lambda_m$ stand for 2-outcome tests while
the rest observables represent 3-outcome tests $(d\ge 3)$. In this
standard base, an arbitrary complete set of LOOs is characterized
by an orthogonal $d^2\times d^2$ real matrix $O$ such that
 \be
 \lambda^o_\mu=\sum_{\nu=1}^{d^2} O_{\mu\nu}\lambda_\nu.
 \ee
Specially $\lambda_\mu^u=u\lambda_\mu u^\dagger$ with $u$ being
unitary is also a complete set of LOOs. Not all LOOs can be
generated by unitary transformations. For example there is no $u$
such that $\lambda_\mu^T=\lambda_\mu^u$, where $\lambda^T_\mu$
denotes the transposition of the standard LOOs.

{\it Entanglement witness.---}Entanglement witness (EW) is an
observable of the composite system that has i) nonnegative
expectation values in all separable states and ii) at least one
negative eigenvalue. We call an observable a candidate of EW if it
satisfies the condition i).

If we choose an arbitrary set of LOOs $\{\lambda^o_\mu\}$ for
system $A$ and the transposition of the standard set
$\{\lambda^T_\mu\}$  for system $B$, then the observable defined
as
 \be\label{ew}
 E_O=I\otimes I-\sum_{\mu=1}^{d^2}\lambda^o_\mu\otimes \lambda^T_\mu
 \ee
is an EW candidate for all orthogonal $O$. This is because in any
product state $\rho=\varrho_1\otimes\varrho_2$ we have
 \begin{eqnarray}
 \tr(\rho
 E_O)&=&1-\sum_{\mu}\avv{\lambda^o_\mu}_1\avv{\lambda^T_\mu}_2\cr
 &\ge&1-\sqrt{\sum_{\mu}\avv{\lambda^o_\mu}_1^2}\sqrt{\sum_{\mu}\avv{\lambda^T_\mu}_2^2}\cr
 &=&1-\sqrt{\tr\varrho_1^2\tr\varrho^2_2}\ge0
 \end{eqnarray}
by using the Cauchy inequality, the orthogonality and completeness
of the LOOs. If the complete set of LOOs is so chosen that the EW
candidate $E_O$ does possess at least one negative eigenvalue then
we obtain an EW. From the proof of the inequality above it is
obvious that instead of $O$ being orthogonal the condition
$OO^T\le 1$ is enough for the construction of an EW candidate. We
shall encounter this kind of EW in the following example.

As an application, we shall construct explicitly an EW for the
$3\times 3$ bound entangled state introduced in Ref.\cite{horo3}.
In the base $\ket{m,n}$ arranged in the ordering $3(m-1)+n$, the
density matrix reads
 \be
 \rho_a=\frac1{1+8a}
 \left(\begin{matrix}a&0&0&0&a&0&0&0&a\cr
               0&a&0&0&0&0&0&0&0\cr
               0&0&a&0&0&0&0&0&0\cr
               0&0&0&a&0&0&0&0&0\cr
               a&0&0&0&a&0&0&0&a\cr
               0&0&0&0&0&a&0&0&0\cr
               0&0&0&0&0&0&\frac{1+a}2&0&\frac{\sqrt{1-a^2}}2\cr
               0&0&0&0&0&0&0&a&0\cr
               a&0&0&0&a&0&\frac{\sqrt{1-a^2}}2&0&\frac{1+a}2\end{matrix}\right).
 \ee
The state $\rho_a$ has PPT while being entangled for all $0<a<1$.
At first we choose special sets of LOOs $\{A_\mu\}$ and
$\{B_\mu\}$ for systems $A$ and $B$ respectively as follows
 \begin{eqnarray}\label{loo}
 A_1&=&\frac1{\sqrt 3}(\lambda_1+\lambda_2+\lambda_3)=B_1,\cr
 A_2&=&\frac1{\sqrt2}(\lambda_{1}-\lambda_2),\quad B_2=\frac1{\sqrt2}(\lambda_{3}-\lambda_1),\cr
 A_3&=&\frac{1+2a}{\sqrt6
 (2+a)}(2\lambda_3-\lambda_1-\lambda_2)-\frac{\sqrt{3(1-a^2)}}{2+a}\lambda_{13}^+,
\cr
 B_3&=&\frac1{\sqrt6}(\lambda_1+\lambda_3-2\lambda_2),\cr
 A_4&=&\frac{1+2a}{2+a}\lambda^+_{13}+\frac{\sqrt{1-a^2}}{\sqrt2(2+a)}(2\lambda_3-\lambda_1-\lambda_2),\cr
 B_4&=&\lambda^+_{13},\quad A_5=\lambda^-_{13}=B_5,\cr
 A_6&=&\lambda^+_{12}=B_6, \quad  A_7=\lambda^-_{12}=B_7,\cr
 A_8&=&\lambda^+_{23}=B_8,\quad A_9=\lambda^-_{23}=B_9.
\end{eqnarray}
A similar choice of local observables has also been used to detect
the entanglement of $\rho_a$ by a local uncertainty relationship
\cite{hofman}. Then we expand the density matrix $\rho_a$ in the
base $\{A_\mu\otimes B_\nu^T\}$ with coefficients
$\rho_{\mu\nu}=\tr(\rho_a A_\mu\otimes B_\nu^T)$. In fact the LOOs
Eq.(\ref{loo}) are designed to make $\sum_\mu\rho_{\mu\mu}=1$. Now
we define a real matrix $M$ with only the following elements
 \be
 M_{\mu\mu}=\frac1{\sqrt{1+n^2}},\quad M_{1\nu}=-M_{\nu 1}=
 \frac{n_\nu}{\sqrt{1+n^2}}
 \ee
nonzero, where the vector defined by
$n_\nu=\rho_{1\nu}-\rho_{\nu1}$ with $\nu=2,\ldots,9$ has a
nonzero norm
 \be
 n^2=\sum_{\nu=2}^{9} n_\nu^2=\frac{(1-a)a^2}{(2+a)(1+8a)^2}.
 \ee
Obviously we have $M^TM\le 1$. So we have an EW candidate
 \be\label{ewa}
 E_a=I\otimes I-\sum_{\mu,\nu=1}^{d^2}M_{\mu\nu}A_\mu\otimes
 B^T_\nu.
 \ee
Its expectation values in separable states are all nonnegative
while its expectation value in the state $\rho_a$ reads
 \be
 \tr(\rho_a E_a)=1-\sqrt{1+n^2}<0
 \ee
for all $0<a<1$. Therefore we obtain explicitly an EW for the
state $\rho_a$. Of course the entanglement of $\rho_a$ can also be
detected by the non-CP map corresponding to the EW $E_a$.

{\it O-reduction criterion.---}To each EW candidate $E_O$ we can
associate a positive map through the Jamio\l kowski isomorphism
\cite{jamio} as follows
 \be
 \mO(\varrho)=\tr_B(I\otimes \varrho^T E_O)=\tr\varrho-\sum_{\mu}
 \avv{\lambda_\mu}_\varrho\lambda^o_\mu\equiv
 \tr\varrho-\varrho^o.
 \ee
Thus we have a separability criterion: if a state $\rho$ of the
composite system is separable then
 \be
 \mO\otimes\mI(\rho)=\tr_A\rho-\rho^{o_A}\ge0,
 \ee
for all orthogonal $O$, where
 \be
 \rho^{o_A}=\sum_{\mu,\nu=1}^{d^2}\av{\lambda_\mu\otimes\lambda_\nu^T}\lambda_\mu^o\otimes\lambda_\nu^T.
 \ee
Not all choices of $O$ will result in a non-CP map. For example if
the orthogonal $O$ is generated by the transposition then the
resulting map is completely positive. If the orthogonal $O$ is
generated by a unitary transformation, then we obtain the
reduction map $\tr\varrho-\varrho$. Thus we refer to the above
separability criterion as {\it O-reduction criterion}. It turns
out that it is exactly the LOOs which cannot be generated by
unitary transformations that are crucial to the detection of bound
entanglement.

As is well known the reduction map is a decomposable non-CP map,
i.e., a composition of the transposition and a completely positive
map. The construction of indecomposable non-CP maps is somewhat
involved, e.g., based on some maximization or minimization
procedures \cite{max} or dependent on some special states
\cite{chen2}. Here we shall see that some simple non-CP maps
induced by the permutation of LOOs are not decomposable, i.e.,
they can be used to detect bound entangled states with PPT.

We consider an arbitrary permutation
$\lambda_\mu\mapsto\lambda_{\sigma(\mu)}$ of the standard LOOs. In
the state $\ket\Phi$ (not normalized) the  expectation value of
the induced EW candidate
 \be
 E_{\sigma}=I\otimes I-\sum_{\mu=1}^{d^2}\lambda_{\sigma(\mu)}\otimes\lambda_\mu^T
 \ee
 reads
$d-\sum_{\mu}\tr(\lambda_{\sigma(\mu)}\lambda_\mu)$, which is
negative if the permutation leaves more than $d+1$ observables
unchanged, i.e., map $\mu\mapsto\sigma(\mu)$ has at least $d+1$
fix points. In this case we obtain an EW and a non-CP map
$\tr\varrho-\varrho^\sigma$.

More specifically we consider a permutation among $d$ observables
$\lambda_m$ in the standard set of LOOs $\lambda_\mu$ while all
other LOOs remain unchanged.  This permutation of LOOs corresponds
to a permutation of the diagonal elements of the density matrix in
the base $\ket n$. Since there are $d^2-d>d+1$ ($d\ge 3$) LOOs
left unchanged, the positive map $\tr\varrho-\varrho^\sigma$ is
not completely positive.

In the following we shall demonstrate the detection power of these
non-CP maps for some entangled states with PPT. This on the other
hand proves that these maps are indecomposable non-CP positive
maps. Let us introduce a state of $d$-level bipartite system as
follows
 \be\label{state}
 \rho=\frac{a_1}d\ket\Phi\bra\Phi+\sum_{k=1,i=2}^{d}\frac{a_i}d
 \lambda_k\otimes\lambda_{k+i-1},
 \ee
where the positive numbers $a_i$ satisfy $\sum_ia_i=1$. In the
case of $d=3$ the state has been discussed in \cite{d3, sym}. It
is not difficult to check that i.) If $a_i\ge a_1$ $(i\ne 1)$ the
state is separable; ii.) If $a_{i+1}a_{d-i+1}\ge a_1^2$ then the
state is a PPT state.

Let us consider $d-1$ cyclic permutations of the diagonal elements
according to rules $\sigma^l(m)=m+l \mod d$ $(l=1,2,\ldots d-1)$.
Applying these permutations to the first subsystem we obtain
 \be
 \rho^{\sigma^l_A}=\rho
  +\frac1d\sum_{k,i=1}^{d}(a_{i-l}-a_i)\lambda_k\otimes\lambda_{k+i-1}
 \ee
By applying the O-reduction criteria, i.e., if the state is
separable then $\tr_A\rho-\rho^{\sigma^l_A}\ge0$, we have to
require $1-a_i\ge (d-1)a_1$ for all $i=1,2,\ldots, d$. Now we
consider a special case where $a_i=a_1$ for $i\ne 2, d$. The
constraints on the separable states from O-reduction criterion
become $a_2\ge a_1$ and $a_d\ge a_{1}$. In this special case the
O-reduction criterion is a necessary and sufficient one for the
separability. As a result we can picture the entanglement of the
state in Eq.(\ref{state}) according to its independent parameters
$a_1$ and $a_2$ in a diagram Fig. 1.
\begin{center}
\begin{figure}
\includegraphics{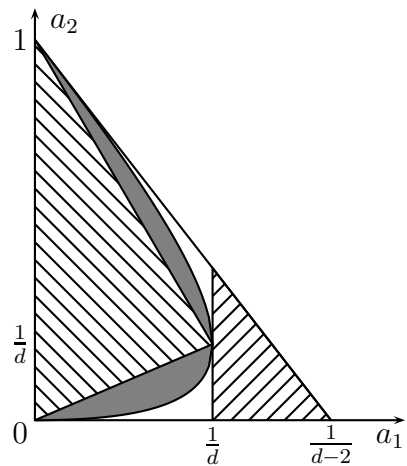}
\caption{The states in the region delineated by the curve
$a_2a_{d}\ge a_1^2$ have PPT, the states within the triangle
inside the curve are separable, and the states within the triangle
outside the curve are free entangled. Therefore the states within
the gray-colored region are bound entangled states.}
 \end{figure}
\end{center}
\vskip -0.8cm

{\it Hermitian correlation matrix.---}One dilemma of entanglement
detection is that the entanglement can be detected only by non-CP
maps but non-CP maps are physically not realizable. Now we realize
the O-reduction criterion by measuring the correlation of LOOs.

Let us take LOOs $\{\lambda^o_\mu\}$  as testing observables for
system $A$ and LOOs $\{\lambda_\mu^T\}$ as testing observables for
system $B$. Their correlations behave differently for separable
states and entangled states. For example the EW candidate $E_O$
imposes a constraint on the correlations of LOOs in a separable
state $\rho$ as
 \be\label{lur}
 \sum_{\mu=1}^{d^2}\av{\lambda^o_\mu\otimes \lambda_\mu^T}\le 1.
 \ee
If the inequality above is maximized over  all possible orthogonal
$O$ we have $\tr\sqrt{TT^T}\le 1$ where $T$ is a $d^2\times d^2$
real correlation matrix with elements
$T_{\mu\nu}=\av{\lambda_\mu\otimes\lambda_\nu^T}$. This is
equivalent to the realignment criterion for separability as soon
we notice that the realigned matrix $\tilde\rho$ defined by
$\bra{m,n}\tilde\rho\ket{k,l}=\bra{m,k}\rho\ket{n,l}$ satisfies
$\tilde\rho=\sum_{\mu\nu}T_{\mu\nu}\ket{\lambda_\mu}\bra{\lambda_\nu}$.
Therefore the realignment criterion is reformulated in terms of
the correlation function on the left-hand side of inequality
Eq.(\ref{lur}).

The inequality Eq.(\ref{lur}) also holds true for the correlations
of more general local observables $\lambda_\mu^o$ where $O$
satisfies $OO^T\le 1$. However, if the inequality Eq.(\ref{lur})
for any orthogonal $O$ fails to identify the entanglement of a
state then the inequality Eq.(\ref{lur}) for any nonorthogonal $O$
fails too since $|\tr(TO)|\le\tr\sqrt{TT^T}$ as long as $OO^T\le
1$.

There are limitations of inequality Eq.(\ref{lur}), e.g.,  there
exist entangled states in $2\times 2$ systems that cannot be
detected by the realignment criterion \cite{chen}. However they
can be detected by the O-reduction criterion because the latter
includes the reduction criterion as a special case. In fact the
realignment criterion is absolutely weaker than the O-reduction
criterion, as can be seen from the fact that
 \be
 \bra\Phi\mO\otimes\mI(\rho)\ket\Phi=1-\tr(TO^T), \quad \forall O.
 \ee
There exist cases in which $|\tr(TO)|\le1$ for all orthogonal $O$
but not all operators $\mO\otimes\mI(\rho)$ are positive.

To achieve a physical detection of entanglement more efficient
than Eq.(\ref{lur}), one has to examine the correlations of local
observables more closely. In stead of a single function of
correlation function as in the left hand side of inequality
Eq.(\ref{lur}), one can build a $d\times d$ Hermitian correlation
matrix $X=\sum_\mu\tr (X\lambda_\mu)\lambda_\mu$ for the
correlations of LOOs $\lambda_\mu^o$ and
$\lambda_\nu^u=u\lambda_\nu u^\dagger$ in state $\rho$ as follows
 \ber
\tr X\lambda_{m}&=&\av{(I-\lambda^o_m)\otimes\lambda_m^u},\\
\tr X\lambda_{mn}^+&=&
\frac{-1}{\sqrt2}\av{\lambda_{mn}^{+o}\otimes\lambda_{mn}^{+u}-
\lambda_{mn}^{-o}\otimes\lambda_{mn}^{-u}},\\
\tr X\lambda_{mn}^-&=&
\frac{-1}{\sqrt2}\av{\lambda_{mn}^{+o}\otimes\lambda_{mn}^{-u}+
\lambda_{mn}^{-o}\otimes\lambda_{mn}^{+u}}.
 \eer
 Now we are able to present
our main result:

{\it Theorem:}  If the state $\rho$ is separable then the
Hermitian correlation matrix is positive, i.e., $X\ge 0$, for
arbitrary unitary $u$ and orthogonal $O$, which is equivalent to
the O-reduction criterion $\mO\otimes\mI(\rho)\ge0$  for all
orthogonal $O$.

Proof. It suffices to prove the second part of the theorem. We
introduce another $d$-level system $C$ as an ancilla and define an
unnormalized state $\ket\Phi_{ABC}=\sum_m\ket{m,m,m}$, then we
have the following relation between the Hermitian correlation
matrix defined above and the O-reduction map
 \be
 X=\tr_{AB}(\mO^T\otimes\mU(\rho)\ket\Phi\bra\Phi_{ABC})
 \ee
where $\mO^T(\varrho)=\tr\varrho-\varrho^{o^T}$ and
$\mU(\sigma)=u^\dagger\sigma u$. It is obvious that if the
O-reduction criterion is satisfied then $X\ge0$. On the other
hand, $X\ge 0$ means that for any real $s_m$ we have
$\sum_{mn}s_ms_nX_{mn}\ge0$. This ensures that
$\mO\otimes\mI(\rho)$ is nonnegative in any pure state  that is
related to  $\sum_ms_m\ket{m,m}$ by local unitary transformations,
which in fact can be any pure state. Thus
$\mO\otimes\mI(\rho)\ge0$ if $X\ge0$ for all unitary $u$ and
orthogonal $O$.

Now let us look at two important special cases. Firstly, the
positivity of the Hermitian correlation matrix $X\ge 0$ means that
$X$ is positive in any state, specially in the state $\sum_m\ket
m$. As a result
$\sum_{\mu}\av{\lambda_\mu^o\otimes\lambda_\mu^u}\le 1$ which is
equivalent to Eq.(\ref{lur}). Secondly, if $d=2$ then $2\times2$
correlation matrix $X\ge0$ is equivalent to the inequality derived
in \cite{yu2}. That is,  $X\ge 0$ is a sufficient and necessary
condition for $2\times 2$ case.

{\it Summary.---}Through local orthogonal observables we have
constructed effective entanglement witnesses and non-CP maps for
states with positive partial transposition. A family of bound
entangled states can be well characterized by the non-CP maps
induced by permutation of local orthogonal observables. Finally
these physically not implementable maps can be realized physically
by measuring the Hermitian correlation matrix, whose negative
eigenvalue (if exists) provides a signature of entanglement.

{\it Acknowledgement.---}Y.S. acknowledges the financial support
of National Natural Science Foundation of China (Grant No.
90303023). N.L. acknowledges the support of the State Education
Ministry of China and the Chinese Academy of Sciences, and
gratefully thanks P.T. Leung for his invaluable support and
encouragement.

\end{document}